\documentclass[journal]{IEEEtran}
\usepackage[none]{hyphenat}
\usepackage{cite}
\usepackage{amsmath}
\usepackage{amssymb,amsfonts}
\usepackage[pdftex]{graphicx}
\usepackage{epstopdf}
\usepackage{textcomp}
\usepackage{xcolor}
\usepackage{bm}
\usepackage{color}
\usepackage{epstopdf}
\usepackage{booktabs}
\allowdisplaybreaks[4]
\begin{document}
\title{Quantum Hamiltonian Identification with Classical Colored Measurement Noise}
\author{Lingyu Tan, Daoyi Dong~\IEEEmembership{Senior Member,~IEEE}, Dewei Li, and Shibei Xue~\IEEEmembership{Member,~IEEE}
\thanks{L.~Tan, S. Xue, and D. W. Li are with Department of Automation, Shanghai Jiao Tong University and Key Laboratory of System Control and Information Processing, Ministry of Education of China, Shanghai, 200240, People's Republic of China (e-mail: \{tanlingyu, dwli, shbxue\}@sjtu.edu.cn).

D. Y. Dong is with the School of Engineering and Information Technology, University of New South Wales, Canberra ACT 2600, Australia. (e-mail: daoyidong@gmail.com).

	
	This work was supported in part by the National Natural Science Foundation of China under Grant 61873162 and 61828303, in part by the Shanghai Pujiang Program under Grant 18PJ1405500.}
}

\maketitle

\begin{abstract}
   In this paper, we present a Hamiltonian identification method for a closed quantum system whose time trace observables are measured with colored measurement noise. The dynamics of the quantum system are described by a Liouville equation which can be converted to a coherence vector representation. Since the measurement process is disturbed by classical colored noise, we introduce an augmented system model to describe the total dynamics, where the classical colored noise is parameterized. Based on the augmented system model as well as the measurement data, we can find a realization of the quantum system with unknown parameters by employing an Eigenstate Realization Algorithm. The unknown parameters can be identified using a transfer-function-based technique. An example of a two-qubit system with colored measurement noise is demonstrated to verify the effectiveness of our method.
\end{abstract}


\section{Introduction}
In recent years, great progress has been achieved on quantum technologies, such as quantum computation~\cite{nielsen2000quantum}, quantum communication~\cite{gisin2007quantum}, and quantum metrology~\cite{giovannetti2011advances}. Relying on the precise models of relevant quantum systems, these quantum-mechanics-based techniques can achieve better performance than classical-mechanics-based counterparts.
 However, under some circumstances, some parameters in these models may not be known, which result in degraded performance. For acquiring these parameters, a fundamental step is identification.
 Classical system identification theory uses some methods, such as wavelet cross spectrum analysis, least squares methods, or maximum-likelihood estimators \cite{Ljung:1986:SIT:21413}, to estimate the system parameters utilizing the input and output data of the system.

 Inspired by classical system identification theory, research on quantum parameter identification is undergoing developments, among which Hamiltonian identification has been widely studied.
Since the Hamiltonian of a quantum system determines the evolution of the quantum state, quantum Hamiltonian identification becomes an important research area, where the identifiability for checking the existence and uniqueness of solutions of quantum Hamiltonian identification is a core issue. An identifiability condition was firstly established for a closed quantum system driven by a laser field, where the populations of all states are measured~\cite{COCV_2007__13_2_378_0}. By relaxing the above measurement condition; i.e., measuring limited observables, an identifiability condition for the same problem was found in~\cite{burgarth2012quantum}.
Alternatively, for a closed many-body quantum system, Sone and Cappellaro ~\cite{PhysRevA.95.022335} analyzed the identifiability for the Hamiltonian identification under a single quantum probe and tested the estimation performance in the presence of Gaussian noises. Also, the identifiability
for linear passive quantum systems coupling to quantum bosonic fields was developed in~\cite{7130587}.

Studies on identifiability for quantum Hamiltonian identification provide theoretical analysis for designing identification algorithms which can be divided into two categories; i.e., time-domain and frequency-domain approaches. In the time-domain approaches, a class of observer-based methods was proposed. Kosut and Rabitz~\cite{KOSUT2002397} used an invariant asymptotic state observer to estimate parameters in Hamiltonian with a gradient algorithm.
 Afterwards, an adaptive observer of exponential convergence was proposed by Bonnabel and Mirrahimi~\cite{ObserverBased} to directly estimate the parameters with Gaussian measurement noises and control. Moreover, quantum tomography technique was applied to quantum Hamiltonian identification. In~\cite{wang2018quantum}, a two-step identification algorithm was presented for a closed quantum system, which is based on the framework of quantum process tomography. Jagadish and Shaji~\cite{JAGADISH2015287} also proposed an algorithm to identify the coupling Hamiltonian between a qubit and its environments using measured data from quantum process tomography. Although the identification algorithm is usually effective, the tomography process is time-consuming. In addition, a system-realization-based method was proposed to identify unknown parameters in the Hamiltonian of a linear passive quantum system~\cite{7130587}.

On the other hand, the frequency-domain approaches were explored for quantum Hamiltonian identification.
Zhang and Sarovar~\cite{PhysRevLett.113.080401} estimated parameters in the Hamiltonian of spin systems using equivalent transfer functions where time trace observables are measured. This method was later extended to open quantum systems~\cite{PhysRevA.91.052121}.
Fourier analysis was also applied to quantum Hamiltonian identification. Cole \emph{et. al.}~\cite{PhysRevA.71.062312} adopted Fourier transform of the measurement of one observable to identify Hamiltonian of a closed two-level quantum system. Schirmer \emph{et. al.}~\cite{1742-6596-107-1-012011} also estimated the Hamiltonian of a two-level quantum system based on Fourier analysis
and Burgarth \emph{et. al.}~\cite{PhysRevA.79.020305} provided the solution for the $N$-level case. In the previous works, the measurement process is assumed to be ideal or disturbed by Gaussian noise, which is convenient for the analysis and design of identification algorithms. However, in practice, we use classical devices in the measurement process which may carry classical colored noise~\cite{zheng2000estimation, ding2009auxiliary}.
Ignoring the impact of colored noise may degrade the performance of the identification algorithms. Nevertheless, it is still an open problem on quantum Hamiltonian identification under the condition that the measurement process is disturbed by classical colored noise.



 In this paper, we propose an augmented system method to identify parameters in the Hamiltonian of a closed quantum system where the data of measured time trace observables carry classical colored noise. We consider a closed finite-level quantum system which can be represented in a coherence vector representation. Correspondingly, when the observable of the system is specified, the dynamics of the system can be described by a reduced equation for the coherence vector. To combine the colored noise into the model for identification, a spectral factorization method is utilized such that an augmented system model for quantum Hamiltonian identification can be obtained.
 Moreover, with the data of the time trace observables, an Eigenstate Realization Algorithm is employed to find a realization of the augmented system. Equalling the transfer function of both the original system with unknown parameters and the realization generated by the measurement data, we can obtain a set of nonlinear algebraic equations for the unknown parameters, which is difficult to be solved analytically. Numerically, these equations can be solved using a PHCpack \cite{journals/toms/Verschelde99}. Finally, we provide an example of a two-qubit system with measurement process disturbed by classical colored noise.

The remainder of this paper is organized as follows.  Section~\ref{systemmodel} describes the model for identification of finite-level quantum systems. In Section~\ref{measureSection}, we develop a colored noise realization which is utilized to augment the original system model. The procedure to obtain identified Hamiltonian is presented in Section~\ref{identification method}. In Section~\ref{example}, the effectiveness of our algorithm is verified in an example of a two-qubit system with classical colored measurement noise. The conclusions are drawn in Section~\ref{conclu}.

\section{Identification model for finite-level quantum system}\label{systemmodel}
In this paper, we consider a closed $N$ dimensional quantum system.
%
The system Hamiltonian $\bm{H}$ satisfies $i\bm{H}\in\mathfrak{su}(N)$, where the Lie algebra $\mathfrak{su}(N)$ can be represented by $N\times N$ traceless skew-Hermitian matrices. The dimension of $\mathfrak{su}(N)$ over $\mathbb{R}$ is $N^2-1$. Therefore we can expand the Hamiltonian as
\begin{equation}\label{hamiltonian}
  \bm{H}=\sum_{m=1}^{N^2-1}a_m\bm{X}_m,
\end{equation}
where $\mathcal{X}=\{i\bm{X}_m,m=1,2,\cdots,N^2-1\}$ is a set of orthogonal bases for $\mathfrak{su}(N)$~\cite{PhysRevLett.113.080401}. The commutation relations for the elements in $\mathcal{X}$ are
\begin{equation}\label{commu}
  [\bm{X}_j,\bm{X}_k]=\sum_{l=1}^{N^2-1}C_{jkl}\bm{X}_l,
\end{equation}
where the commutator $[\cdot,\cdot]$ is calculated as $[\bm{X},\bm{Y}]=\bm{XY}-\bm{YX}$ for two operators $\bm{X}$ and $\bm{Y}$ and $C_{jkl}$ are antisymmetric constants with respect to the interchange of any pair of indices~\cite{cartier1966quantum}.  This property indicates that $C_{jkl}$ equals to zero with any two identical indices.

%
The Hamiltonian determines the dynamics of the density matrix $\bm{\rho}(t)$ of the system as
\begin{equation}\label{Liouville}
  \bm{\dot{\rho}}(t)=-i[\bm{H},\bm{\rho}(t)].
\end{equation}
which is the so-called Liouville equation~\cite{smith1991introduction}. The density matrix $\bm{\rho}$ describes the probability distribution of the system states, which is an $N\times N$ Hermitian and positive semi-definite matrix with $\text{tr}(\bm{\rho})=1$. We have assumed $\hbar=1$.

The Liouville equation (\ref{Liouville}) can be transformed into a coherence vector representation~\cite{xueCST2018} which is convenient for the design of the identification algorithm. In this representation, the state of the system is alternatively described by a coherence vector $x=[x_1,\cdots,x_{N^2-1}]^T$ where $x_j$ is the expectation value of $\bm{X}_j$; i.e., $x_j=\text{tr}(\bm{X}_j\bm{\rho})$. The corresponding dynamical equation is thus written as
%
\begin{equation}\label{dynamics}
\begin{split}
  &\dot{x}_j(t)=i\sum_{l=1}^{N^2-1}(\sum_{m=1}^{N^2-1}a_mC_{mjl})x_l(t),\\ &j=1,2,\cdots,N^2-1.
  \end{split}
\end{equation}

%
%

To observe the quantum system, we can choose $L$ observables ${\bm{O}_1,\bm{O}_2,\cdots,\bm{O}_L}$ and their expectations can be taken as the outputs of the system; i.e.,
\begin{equation}\label{output}
  \bm{y}(t)=\left[
                \begin{array}{cccc}
                 \bm{y}_1(t), &\bm{y}_2(t), & \cdots,&\bm{y}_L(t) \\
                \end{array}
              \right]^T\nonumber\\
\end{equation}
where $\bm{y}_i(t)=\langle\bm{O}_i\rangle={\rm tr}[\bm{O}_i\bm{\rho}]$.
An observable $\bm{O}_i$ can be expanded in terms of the bases in $\mathcal{X}$ as $\bm{O}_i=\sum_{j}o_j^{(i)}\bm{X}_j$ where the corresponding bases $\mathcal{M}=\{\bm{X}_{\bm{\mu}_1},\bm{X}_{\bm{\mu}_2},\cdots,\bm{X}_{\bm{\mu}_p}\}$ span a minimal space containing the $L$ observables. Here, $\bm{\mu}_i$ denotes the indices for the corresponding bases and the number of $\bm{\mu}_i$ is $p$.

With respect to the bases in $\mathcal{M}$,
an accessible set of $\mathcal{M}$ can be generated using a filtration process~\cite{PhysRevLett.113.080401}.
Denoting $\mathcal{F}_0=\mathcal{M}$,  an iterative procedure can be calculated $\mathcal{F}_i=\mathcal{F}_{i-1}\cup[\mathcal{F}_{i-1}, \mathcal{X}]$, where $[\mathcal{F}_{i-1}, \mathcal{X}]=\{\bm{X}_j|\text{tr}(\bm{X}_j^{\dag}[g,h])\neq0, g\in\mathcal{F}_{i-1},h\in\mathcal{X}\}$, until $\mathcal{F}_i$ saturates.
Supposing the final set is $\mathcal{F}=\{\bm{X}_{\bm{\mu}_1},\bm{X}_{\bm{\mu}_2},\cdots,\bm{X}_{\bm{\mu}_K}\}$ with a size $K$, the corresponding reduced coherence vector is $\bm{x}=[x_{\bm{\mu}_1},x_{\bm{\mu}_2},\cdots,x_{\bm{\mu}_p},\cdots,x_{\bm{\mu}_K}]^T$. 
Therefore, the corresponding reduced dynamical equation for the reduced coherence vector can be written as
\begin{equation}\label{system_reali}
  \begin{split}
  &\dot{\bm{x}}(t)=\bm{Ax}(t), \bm{x}(0)=\bm{x}_0\\
  &\bm{y}(t)=\bm{Cx}(t),\\
  & \bm{A}\in\mathbb{R}^{K\times K}, \bm{C}\in \mathbb{R}^{L\times K},
  \end{split}
\end{equation}
where $\bm{x}_0$ is the initial state, $\bm{A}_{jl}=-i\sum_{m=1}^{N^2-1}a_mC_{m\bm{\mu}_j\bm{\mu}_l}$ and $\bm{C}$ is configured such that $\bm{y}(t)$ are expectation values of our measured observables.
The equation (\ref{system_reali}) affords the basic model for the identification problem.

\section{Augmented Model For Output Disturbed By Classical Colored Noise}\label{measureSection}
In existing Hamiltonian identification studies, ideal case or Gaussian noise disturbing measurement results have been considered. However, in practice, the measurement results may be polluted by classical colored noise arising from measurement devices. Hence, it is necessary to introduce the classical colored noise into the dynamics and thus we obtain a complete model for the purpose of Hamiltonian identification.
%

\subsection{Classical Colored Noise Model}\label{noisemodel}
In general, classical colored noise can be characterized by a shaped power spectral density (PSD) $S(\omega)$ describing the signal power distribution over all the frequency components~\cite{norton_karczub_2003}. Since the PSD and autocorrelation $R(t)$ form a Fourier transform pair, the PSD can be obtained by
\begin{equation}\label{power}
  S(\omega)=\int_{-\infty}^{\infty}e^{-i\omega\tau}R(\tau)\mathrm{d}\tau.
\end{equation}
%
The spectral factorization theorem tells that a positive, rational and strictly proper PSD $S(\omega)$ can be factorized as
\begin{equation}\label{fact}
S(\omega)=\Gamma(s)\Gamma^{T}(-s)|_{s=i\omega},
\end{equation}
where $\Gamma(s)$ is a causal transfer function which results from the internal dynamics of noises. Here, $\Gamma(s)$ characterizes a mapping of a white noise input $\eta(s)=\mathcal{L}[\eta(t)]$ to a colored noise output $v(s)=\mathcal{L}[v(t)]$. The operator $\mathcal{L}[\cdot]$ is Laplace transform and $\eta(t)$ and $v(t)$ are the white noise input and the colored noise in the time domain, respectively. Note that the power spectral density of $v(t)$ is $S(\omega)$ which will reduce to a flat one when the output noise is white.
Also, for an irrational power spectral density, we can find its rational approximants using Pad{\'e} approximation \cite{baker1996padé} and then a transfer function can be obtained by the factorization.

For a given transfer function $\Gamma(s)$, it is easy to construct a corresponding minimal realization \cite{callier2012linear}. For the single-input-single-output (SISO) transfer function
\begin{equation}\label{transfer}
  \Gamma(s)= \frac{\beta_1s^{n-1}+\beta_2s^{n-2}+\cdots+\beta_{n-1}s+\beta_n}{s^n+\alpha_1s^{n-1}+\cdots+\alpha_{n-1}s+\alpha_n}
\end{equation}
which is strictly proper, we can write its realization in a controllable canonical form as
\begin{equation}\label{reali}
\begin{split}
  &\dot{\bm{\xi}}(t)=\bm{E}\bm{\xi}(t)+\bm{F}\eta(t),\\
  &v(t)=\bm{G}\bm{\xi}(t),
\end{split}
\end{equation}
with
\begin{equation}\label{controllable}
\begin{split}
  &\bm{E}=\left[\begin{array}{ccccc}
                   0 & 1 & 0 & \cdots & 0 \\
                   0 & 0 & 1 & \cdots & 0 \\
                   \vdots & \vdots & \vdots & \ddots & \vdots \\
                   0 & 0 & 0 & \cdots & 1 \\
                   -\alpha_n & -\alpha_{n-1} & -\alpha_{n-2} & \cdots & -\alpha_1
                 \end{array}\right],\\
  &\bm{F}=\left[0\quad 0\quad \cdots\quad 0\quad 1\right]^T,\\
  &\bm{G}=[\beta_n\quad \beta_{n-1}\quad \beta_{n-2}\quad\cdots \quad \beta_1],
\end{split}
\end{equation}
where $\bm{\xi}\in\mathbb{R}^n$ can be considered as the internal state vector of the noise with an initial state $\bm{\xi}(0)=\bm{\xi}_0$.
It is straightforward to obtain the dynamics of the expectation $\bm{\bar{\xi}}$ of the internal mode $\bm{{\xi}}$ as
\begin{equation}\label{noiseexpect}
\begin{split}
  &\bm{\dot{\bar{\xi}}}(t)=\bm{E}\bm{\bar{\xi}}(t), \bm{\bar{\xi}}(0)=\bm{\bar{\xi}}_0, \\
  &\bar{v}(t)=\bm{G}\bm{\bar{\xi}}(t),
\end{split}
\end{equation}
where $\bar{v}(t)$ is the expectation of the colored noise $v(t)$ and the input term vanishes since $\bar{\eta}(t)=0$. 
\subsection{Augmented System Model}\label{AugmentedSystemModel}
%
%

   We consider that the classical colored noise $v(t)$ is additive and we assume that the $i$-th output $\bm{y}_i(t)$ of the quantum system; i.e., the time trace of the $i$-th observable $\bm{O}_i$, is disturbed by the expectation value of the classical colored noise. Hence, the polluted output can be expressed as
   \begin{equation}\label{assumption}
     \tilde{\bm{y}}_i(t)=\bm{y}_i(t)+\bar{v}(t),
   \end{equation}
    where $\bm{y}_i(t)$ is the original quantum output for the $i$-th observable and $\tilde{\bm{y}}_i(t)$ is the polluted one.
Hence, the polluted output of the system can be written as
\begin{equation}\label{noisyoutput}
\begin{split}
  \bm{\tilde{y}}(t)=&\left[\begin{array}{c}
                           \bm{y}_1(t)+\bar{v}(t)\\
                           \bm{y}_2(t)+\bar{v}(t)\\
                           \vdots\\
                           \bm{y}_L(t)+\bar{v}(t)
                         \end{array}\right]
           =\bm{Cx}(t)+\left[\begin{array}{c}
           1\\
           1\\
           \vdots\\
           1
           \end{array}\right]_{L\times 1}\bar{v}(t).
\end{split}
\end{equation}

Further, taking the expression of $\bar v(t)$ in (\ref{noiseexpect}) into (\ref{noisyoutput}) and denoting a new state vector $\bm{\check{x}}$ as $\bm{\check{x}}=\left[\begin{array}{cc}
               \bm{x}(t)^T&
               \bar{\bm{\xi}}(t)^T
             \end{array}\right]^T$, we can combine the original system with the noise model as
\begin{equation}\label{simplemodel}
\begin{split}
   &\dot{\bm{\check{x}}}(t)=\bm{\check{A}}\bm{\check{x}}(t), \bm{\check{x}}(0)=\bm{\check{x}}_0,\\
   &\bm{\tilde{y}}(t)=\bm{\check{C}}\bm{\check{x}}(t),
\end{split}
\end{equation}
with
\begin{eqnarray}\label{augmentedmodle}
    \bm{\check{A}}& =& \left[\begin{array}{cc}
                     \bm{A} & 0 \\
                     0 & \bm{E}
             \end{array}\right],\\
            \bm{\check{C}}& =&\left[
                                \begin{array}{cc}
                                \bm{C} & \bm{G}_L \\
                                \end{array}
                              \right],\\
    \bm{G}_L&=& \left[\begin{array}{c}
                            \bm{G} \\
                            \bm{G} \\
                            \vdots \\
                            \bm{G}
                          \end{array}\right]_{L\times n},
\end{eqnarray}
where $\bm{\check{x}}\in\mathbb{R}^{\check{n}}, \bm{\check{A}}\in\mathbb{R}^{\check{n}\times\check{n}},
                          \bm{\check{C}}\in\mathbb{R}^{L\times\check{n}}$ and the initial state of the augmented state is $\bm{\check{x}}_0=\left[\begin{array}{cc}
                                     \bm{x}_0^T &
                                     \bar{\bm{\xi}}_0^T
                                     \end{array}\right]^T.$ The order of the augmented model $\check{n}$ satisfies $\check{n}=K+n$ where $K$ and $n$ are the orders of the quantum system model and the colored noise realization, respectively.

Now, we obtain an augmented system (\ref{simplemodel}) of the finite-level quantum system whose output is disturbed by classical colored noise. A similar model can be found in the design of Kalman filter under classical colored measurement noise \cite{gomez2011kalman}.
Note that both the dynamics of the quantum system and the classical colored noise contribute to the polluted outputs $\bm{\tilde{y}}(t)$. However, it is difficult to distinguish the unpolluted quantum output $\bm{{y}}(t)$ from the noise directly. A possible method is in demand for extracting the quantum information of the original system from the polluted output.


\section{Hamiltonian Identification with the Time Trace Observables Polluted by Classical Colored Noise}\label{identification method}
\subsection{Problem Statement}

In Section~\ref{measureSection}, we have introduced the dynamics of the coherence vector for the finite-level quantum system and considered the measurement process disturbed by classical colored noise which is represented by a linear system realization for a given spectrum density $S(\omega)$. Consequently, we have obtained a parameterized augmented system model for describing the total dynamics of the coherence vector and the internal modes of the noise. We aim to identify the unknown coefficients in $\{a_m\in \mathbb{R}, m=1,2,\cdots,M\}$ for a quantum system with the Hamiltonian (\ref{hamiltonian}) using the augmented model.
Hence, we state our Hamiltonian identification problem as follows.

Given the structure of the augmented system model (\ref{simplemodel}) for the closed quantum system with an initial state $\bm{\check{x}}_0$, our identification problem is to estimate the unknown parameters $\{a_m\in \mathbb{R}, m=1,2,\cdots,M\}$ in the model utilizing the disturbed measurements $\bm{\tilde{y}}(t)$ of the time traces of the observables for the underlying system.

\subsection{Measurement Process}\label{Measuring process}
To access the information of the quantum system, we measure the time trace observables (\ref{output}). We assume that we have a large number of copies of the system prepared in an identical initial state. Thus we can measure an observable on many identical systems at a time and then obtain the expectation value of the observable. We call a set of identical systems we measure as an ensemble. We measure in this way because in quantum mechanics quantum measurement for any observable will generally change the system state. Moreover, since we cannot obtain measurement results for the same system at different time instants, we need to make measurements on different ensembles for different instants.

\begin{figure}[htbp]
	\centerline{\includegraphics[width=3in]{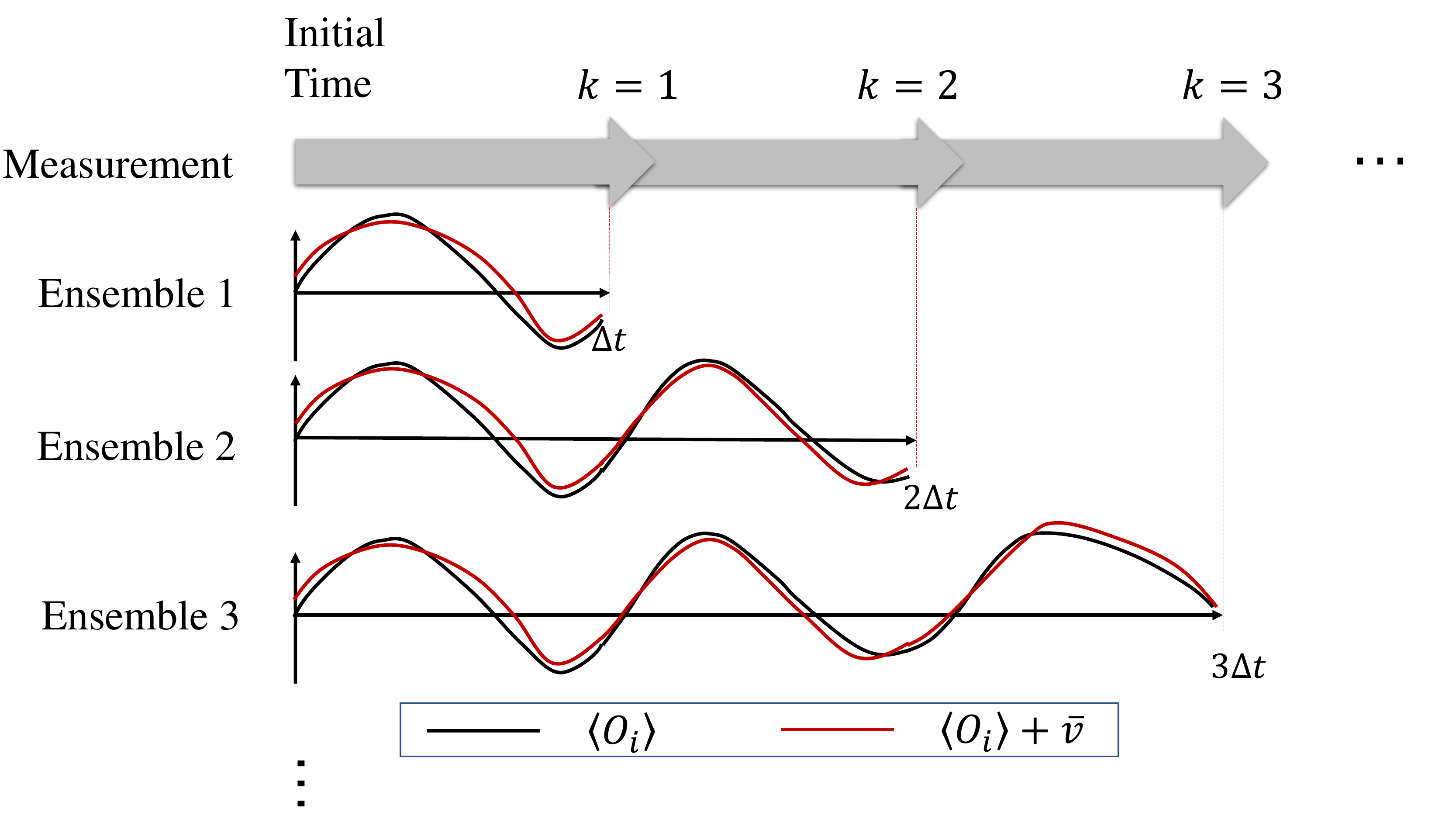}}
	\caption{The process of measuring the time trace of observable $\bm{O}_i$.}
	\label{measure}
\end{figure}

Concretely speaking, we sample the observables with an equal interval $\Delta t$.
  Denoting the measured value of $\bm{O}_i$ of the $j$-th copy at a time instant $k\Delta t$ as $y_i^{(j)}(k)$, the measured value of $\bm{O}_i$ accompanied by colored noise $v^{(j)}(k)$ can be expressed as $\tilde{y}_i^{(j)}(k)=y_i^{(j)}(k)+v^{(j)}(k)$. Here, we have written $\{y_i(k\Delta t)\}$ as $\{y_i(k)\}$ for simplicity. After measuring many copies at different time instants, we can average over the measurement results and obtain (\ref{assumption}). The measurement process is shown in Fig. \ref{measure}.


Note that when we consider multi-observable time traces, commutative observables can be measured simultaneously. However, due to the uncertainty principle \cite{PhysRev.34.163}, measurements of non-commutative observables should be carried out on different ensembles.


\subsection{Identification Algorithm for Measurement Data with Classical Colored Noise}
To link the discrete measurement data to the system model of the augmented system model (\ref{simplemodel}), we discretize (\ref{simplemodel}) for a given sampling interval $\Delta t$ and thus we have
\begin{eqnarray}\label{discrete}
\bm{\check{x}}(k+1)&=&\bm{\check{A}}_d\bm{\check{x}}(k),\nonumber\\
\bm{\tilde{y}}(k)&=&\bm{\check{C}}\bm{\check{x}}(k),
\end{eqnarray}
with $\bm{\check{x}}(0)=\bm{\check{x}}_0$, where $\bm{\check{A}}_d=e^{\bm{\check{A}}\Delta t} \in\mathbb{R}^{\check{n}\times\check{n}},$ $\bm{\check{C}}\in\mathbb{R}^{L\times\check{n}}$, and $k$ stands for the $k$-th time step. Hence, the initial state response of (\ref{discrete}) can be written as
\begin{equation}\label{Markov}
\bm{\tilde{y}}(k)=\bm{\check{C}}\bm{\check{A}}_d^k\bm{\check{x}}_0.
\end{equation}

It is difficult to solve $\{a_m\}$ from (\ref{Markov}) since it is transcendental in $\{a_m\}$ and the system dimension of $\check{n}$ is unknown.
However, these difficulties can be overcome by combining a system realization method~\cite{PhysRevLett.113.080401} and an Eigenstate Realization algorithm (ERA)~\cite{juang1985eigensystem}. Using the ERA, the dimension $\check{n}$ can be determined. In addition, a system realization can be constructed based on the measurement data such that the unknown parameters can be obtained by solving a set of nonlinear equations arising from the the equivalence between the transfer functions of the original system and the realization. It should be mentioned that we calculate the realization using the measurement data disturbed by classical colored noise naturally which results from both the dynamics of quantum system and the noise.

The ERA begins with a generalized Hankel matrix
\begin{equation}\label{Hamkel}
\small
\begin{split}
  &\bm{H}_{rs}(k)=\\
  &\left[\begin{array}{cccc}
                         \bm{\tilde{y}}(k) & \bm{\tilde{y}}(k+t_1) & \cdots & \bm{\tilde{y}}(k+t_{s-1}) \\
                         \bm{\tilde{y}}(k+j_1) & \bm{\tilde{y}}(k+j_1+t_2) & \cdots & \bm{\tilde{y}}(k+j_1+t_{s-1}) \\
                         \vdots & \vdots &  & \vdots \\
                         \bm{\tilde{y}}(k+j_{r-1}) & \bm{\tilde{y}}(k+j_{r-1}+t_2) & \cdots & \bm{\tilde{y}}(k+j_{r-1}+t_{s-1})
                       \end{array}\right]
\end{split}
\end{equation}
which is constructed by the measurement result. Its dimension is $rL\times s$ with two integers $r$ and $s$. To determine $\check{n}$ accurately, it is good to choose a sufficient number of measurement results; i.e., two large integers $r$ and $s$ are preferred.


We choose the measurement results from the initial time; i.e., $k=0$, and then we can have a singular value decomposition of $\bm{H}_{rs}(0)$ as
\begin{equation}\label{svd}
\bm{H}_{rs}(0)=\bm{P}\left[\begin{array}{cc}\bm{D}&\bm{0}\\\bm{0}&\bm{0}\end{array}\right]\bm{Q}^T=[\bm{P}_1\quad \bm{P}_2]\left[\begin{array}{cc}\bm{D}&\bm{0}\\\bm{0}&\bm{0}\end{array}\right]\left[\begin{array}{c}\bm{Q_1}^T\\\bm{Q}_2^T\end{array}\right],
\end{equation}
where $\bm{P}\in\mathbb{R}^{rL\times rL}, \bm{Q}\in\mathbb{R}^{s\times s}$ are unitary matrices, and they are partitioned into $\bm{P}_1, \bm{P}_2$, and $\bm{Q}_1, \bm{Q}_2$ with respect to the dimension of the diagonal square matrix of $\bm{D}$.  Since the diagonal elements of $\bm{D}$ are positive singular values of $\bm{H}_{rs}(0)$, its dimension is determined by the number of the singular values.
%
%
%

Using the identity matrix $\bm{I}$ with a subscript indicating its dimension, we define matrix $\bm{E}^T_L=[\bm{I}_L,\bm{0}_L,\cdots,\bm{0}_L]_{L\times rL}$ and $\bm{e}_1$ is the first column of $\bm{I}_s$.
According to \cite{juang1985eigensystem}, we can establish a numerical realization
\begin{equation}\label{ERA}
\begin{split}
   &\bm{\hat{x}}(k+1)=\bm{\hat{A}}_d\bm{\hat{x}}(k), \bm{\hat{x}}(0)=\bm{\hat{x}}_0,\\
   &\bm{\tilde{y}}(k)=\bm{\hat{C}}\bm{\hat{x}}(k),\\
\end{split}
\end{equation}
with $\bm{\hat{A}}_d=\bm{D}^{-1/2}\bm{P}_1^T\bm{H}_{rs}(1)\bm{Q}_1\bm{D}^{-1/2}$, $\bm{\hat{C}}=\bm{E}^T_L\bm{P}_1\bm{D}^{1/2}$, and $\bm{\hat{x}}_0=\bm{D}^{1/2}\bm{Q}_1^T\bm{e}_1$ \cite{juang1985eigensystem}. It is clear that the order of this realization is exactly the dimension of $\bm{D}$. Further, because of the equivalence between  system models (\ref{discrete}) and (\ref{ERA}), the dimension $\check{n}$ of uncertain model (\ref{discrete}) is expected to be the same as that of (\ref{ERA}) or $\bm{D}$.
Then letting $\bm{\hat{A}}=\log \bm{\hat{A}}_d/\Delta t$, the pair $(\bm{\hat{A}}, \bm{\hat{C}}, \bm{\hat{x}}_0)$ formulates a continuous-time realization describing dynamics of both quantum system and colored noise.

 Thus far, we have completed the process of developing a numerical time-continuous realization from measured data and also determined the system dimension $\check{n}$. Moreover, we have built up an augmented model $(\bm{\check{A}}, \bm{\check{C}}, \bm{\check{x}}_0)$ in (\ref{simplemodel}). The corresponding transfer functions from the initial states to the outputs of the two models should be equal \cite{callier2012linear}; that is,
\begin{equation}\label{transfereq}
 \bm{\check{C}}(s\bm{I}_{\check{n}}-\bm{\check{A}})^{-1}\bm{\check{x}}_0
 =\bm{\hat{C}}(s\bm{I}_{\check{n}}-\bm{\hat{A}})^{-1}\bm{\hat{x}}_0.
\end{equation}
The left-hand side contains the parameters to be estimated, while the right-hand side is completely determined by measured data. To solve $\{a_m\}$ from this equation, we just need to equal the coefficients of $s$ in all orders of the both sides. Firstly, the left-hand side can be simplified as $Q(s)/P(s)$ \cite{callier2012linear}, where
\begin{equation}\label{PQ}
\begin{split}
  &P(s)=\det(s\bm{I}_{\check{n}}-\bm{\check{A}})\\
  &Q(s)=\det\big(s\left[\begin{array}{cc}
                     \bm{I}_{\check{n}} & \bm{0} \\
                     \bm{0} & \bm{0}
                   \end{array}\right]-\left[\begin{array}{cc}
                                              \bm{\check{A}} & \bm{\check{x}}_0 \\
                                              \bm{\check{C}} & \bm{0}
                                            \end{array}\right]\big).
\end{split}
\end{equation}
In fact, the coefficients of $s$ in different orders in $P(s),Q(s)$ are polynomials of the unknown parameters $\{a_m\}$, and the corresponding coefficients in the right hand are numbers obtained through experiment. Next, what we need to do is to solve these polynomial equations for unknown Hamiltonian parameters $\{a_m\}$. Since these equations are often nonlinear or in high order, professional numerical tools such as PHCpack \cite{journals/toms/Verschelde99} can be used to obtain final results. To this end, we have introduced the whole procedure to identify unknown parameters in the Hamiltonian under disturbed measurements with classical colored noise . This procedure can be applied to finite-dimensional closed quantum system with measurements disturbed by classical colored noise as long as the noise can be represented by linear system models though spectral factorization of its PSD.

\section{An Example for A Two-qubit System}\label{example}
 In this section, we consider a two-qubit system whose Hamiltonian is written as
\begin{equation}\label{H2}
  \bm{H}=\sum_{\alpha=1}^{2}\frac{\omega_\alpha}{2}\bm{\sigma}_z^\alpha+\delta_1(\bm{\sigma}_+^1\bm{\sigma}_-^2+\bm{\sigma}_-^1\bm{\sigma}_+^2),
\end{equation}
with Pauli matrices
\begin{equation}\label{sigmaz}
  \bm{\sigma}_z=\left[\begin{array}{cc}
                        1 & 0 \\
                        0 & -1
                      \end{array}\right],\bm{\sigma}_x=\left[\begin{array}{cc}
                        0 & 1 \\
                        1 & 0
                      \end{array}\right],\bm{\sigma}_y=\left[\begin{array}{cc}
                         0& -i \\
                        i & 0
                      \end{array}\right],
\end{equation}
and the corresponding ladder operators
\begin{equation}\label{ladder}
  \bm{\sigma}_+=\left[\begin{array}{cc}
                        0 & 1 \\
                        0 & 0
                      \end{array}\right], \bm{\sigma}_-=\left[\begin{array}{cc}
                        0 & 0 \\
                        1 & 0
                      \end{array}\right].
\end{equation}
Here, the superscripts in the Hamiltonian label the qubits.
For the identification aim, we measure the local observable $\bm{\sigma}_x^1$ of the first qubit and thus with the observable-induced accessible set and the Hamiltonian, a dynamical equation of the coherence vector of the two-qubit system can be written as
\begin{eqnarray}\label{realization2}
  \left[\begin{array}{c}
          \dot{x}_1(t) \\
          \dot{x}_2(t) \\
          \dot{x}_3(t) \\
          \dot{x}_4(t)
        \end{array}\right]&=&\left[\begin{array}{cccc}
                                   0 & -\omega_1 & 0 & \delta_1 \\
                                   \omega_1 & 0 & -\delta_1 & 0 \\
                                   0 & \delta_1 & 0 & -\omega_2 \\
                                   -\delta_1 & 0 & \omega_2 & 0
                                 \end{array}\right]\left[\begin{array}{c}
                                                           x_1(t) \\
                                                           x_2(t) \\
                                                           x_3(t) \\
                                                           x_4(t)
                                                         \end{array}\right]\nonumber\\
  y(t)&=&\left[
           \begin{array}{cccc}
             1 & 0 & 0 & 0 \\
           \end{array}
         \right]
  \left[\begin{array}{c}
                                                           x_1(t) \\
                                                           x_2(t) \\
                                                           x_3(t) \\
                                                           x_4(t)
                                                         \end{array}\right],
\end{eqnarray}
where $x_1(t)=\langle\bm{\sigma}_x^1(t)\rangle$, $x_2(t)=\langle\bm{\sigma}_y^1(t)\rangle$, $x_3(t)=\langle\bm{\sigma}_z^1\bm{\sigma}_x^2(t)\rangle$ and $x_4(t)=\langle\bm{\sigma}_z^1\bm{\sigma}_y^2(t)\rangle$. The initial state is set as $[x_1(0)\quad x_2(0)\quad x_3(0)\quad x_4(0)]^T=\left[
           \begin{array}{cccc}
             0& 1 & 0 & 0 \\
           \end{array}
         \right]^T$.
To simulate the real dynamics of the quantum system, we set the real parameters as $\omega_1=1.3~\text{GHz}, \omega_2=2.4~\text{GHz}$ and $\delta_1=4.3~\text{GHz}$.

As for the classical colored noise $v(t)$ added in measurement process, we assume that its power spectrum density is expressed as
\begin{equation}\label{psd}
  S(\omega)=\frac{10^{12}\omega^2+4\times 10^{26}}{\omega^4-3.999\times 10^{13}\omega^2+4\times 10^{26}}.
\end{equation}
Factorizing the spectrum $S(\omega)$, we can obtain a transfer function $\Gamma(s)$ as below
\begin{equation}\label{trans}
  \Gamma(s)=\frac{10^6s-2\times 10^{13}}{s^2+10^5s+2\times 10^{13}}.
\end{equation}
A realization in a controllable canonical form can be found as
\begin{equation}\label{noise model}
  \begin{split}
  & \dot{\bm{\xi}}(t)=\left[\begin{array}{cc}
                    0 & 1 \\
                    -2\times 10^{13} & -10^5
                  \end{array}\right]\bm{\xi}(t)+\left[\begin{array}{c}
                                                      0 \\
                                                      1
                                                    \end{array}\right]\eta(t),\\
  & v(t)=[-2\times 10^{13}\quad 10^6]\bm{\xi}(t).
  \end{split}
\end{equation}
with a two dimensional internal mode $\bm{\xi}(t)$.

\begin{figure}[htbp]
\centerline{\includegraphics[width=3in]{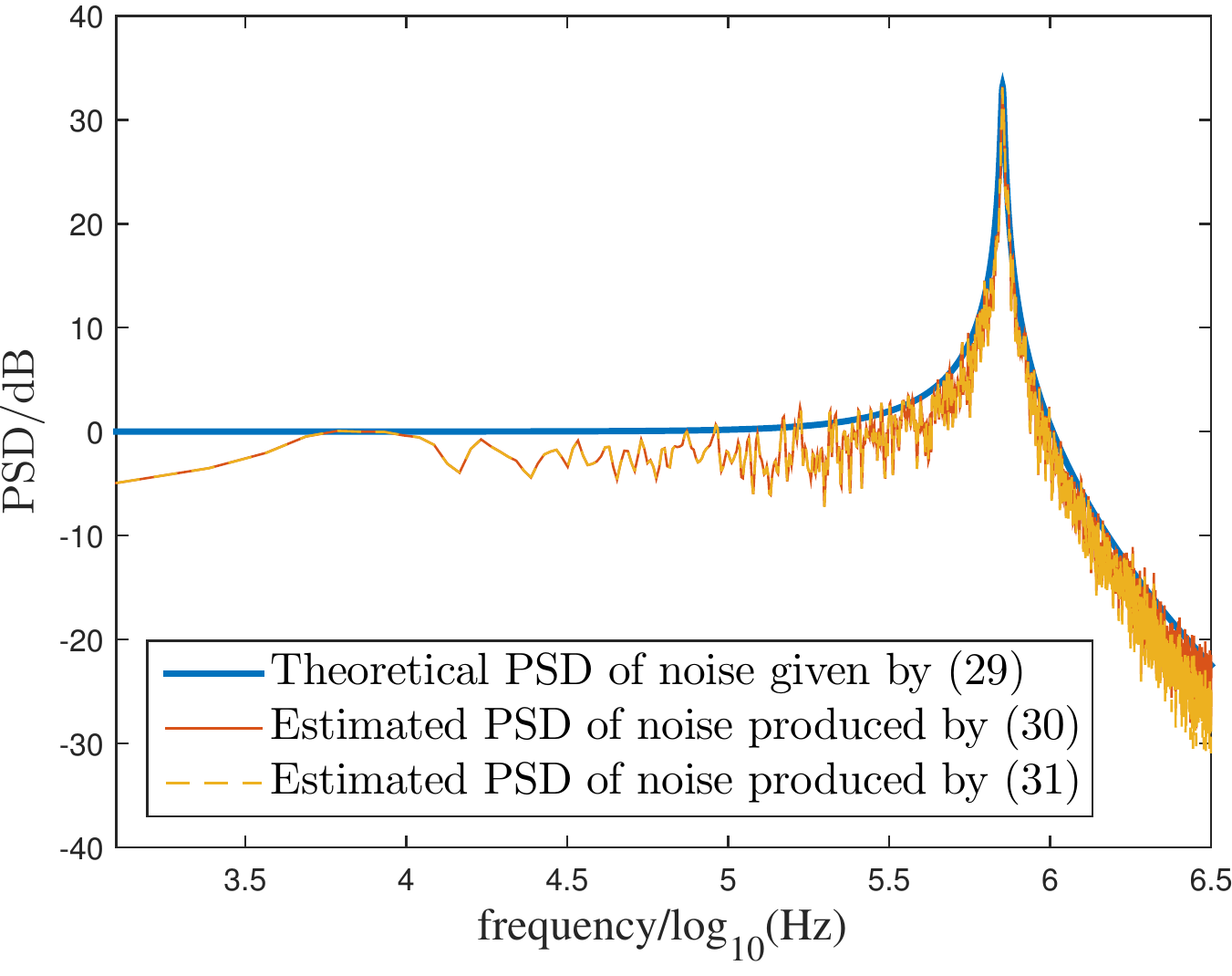}}
\caption{Comparison between the theoretical colored noise PSD and the estimated PSDs of two colored noise signals produced from the transfer function (\ref{trans}) and the realization (\ref{noise model}), respectively.
}
\label{noise_PSD}
\end{figure}

We verify the validity of this noise realization (\ref{noise model}) by checking its power spectral density. First, imposing a white noise signal $\eta(t)$ on both the transfer function (\ref{trans}) and the noise realization (\ref{noise model}) with an arbitrary initial state, we obtain two output colored noise signals. Then we estimate the PSDs of the two signals using Welch's overlapped segment averaging estimator \cite{welch1967use}. The two estimated PSD curves  are compared with the theoretical PSD (\ref{psd}) as shown in Fig. \ref{noise_PSD}. It can be seen that the three curves are of a similar tendency. However, we witness a disparity due to the imperfect white noise and the error for estimating the PSD. Therefore, in simulation, we can produce the expectation of colored noises using the realization (\ref{noise model}) with an arbitrary initial state.

The parameters to be identified are $\omega_1, \omega_2$ and $\delta_1$ as well as the matrices and the state in (\ref{noiseexpect})
for the classical colored noise. Before we analyze the measurement data, we may not know the order $n$ for the colored noise model. However, with the noise model, we can still write an augmented system model for the parameter identification as
\begin{equation}\label{system}
  \begin{split}
       & \dot{\bm{\check{x}}}(t)=\bm{\check{A}}\bm{\check{x}}(t), \\
       & \tilde{\bm{y}}(t)=\bm{\check{C}}\bm{\check{x}}(t)
  \end{split}
\end{equation}
with
\begin{equation}\label{AC}
\begin{split}
  &\bm{\check{A}}=\left[\begin{array}{ccccc}
                                   0 & -\omega_1 & 0 & \delta_1& \\
                                   \omega_1 & 0 & -\delta_1 & 0 &\\
                                   0 & \delta_1 & 0 & -\omega_2 &\\
                                   -\delta_1 & 0 & \omega_2 & 0&\\
                                   & & & & \bm{E}
                                 \end{array}\right],\\
   & \bm{\check{C}}= [1\quad 0\quad 0\quad 0\quad \bm{G}],\\
   & \bm{\check{x}}(0)=[0,1,0,0,\bm{\xi}(0)^T]^T,
   \end{split}
\end{equation}
where the augmented state vector is $\bm{\check{x}}(t)=[x_1(t),x_2(t),x_3(t),\bar{\bm{\xi}}(t)^T]^T$.
 Note that for the classical colored noise, there exist many realizations which are equivalent and related by a similarity transformation.
 Therefore, we can just fix the part elements of the matrices in the noise realization to reduce the parameters to be identified. In our simulation, we assume the output vector $\bm{G}=[1,1,\cdots,1]_{1\times n}$.

\begin{figure}[h]
	\centerline{\includegraphics[width=3in]{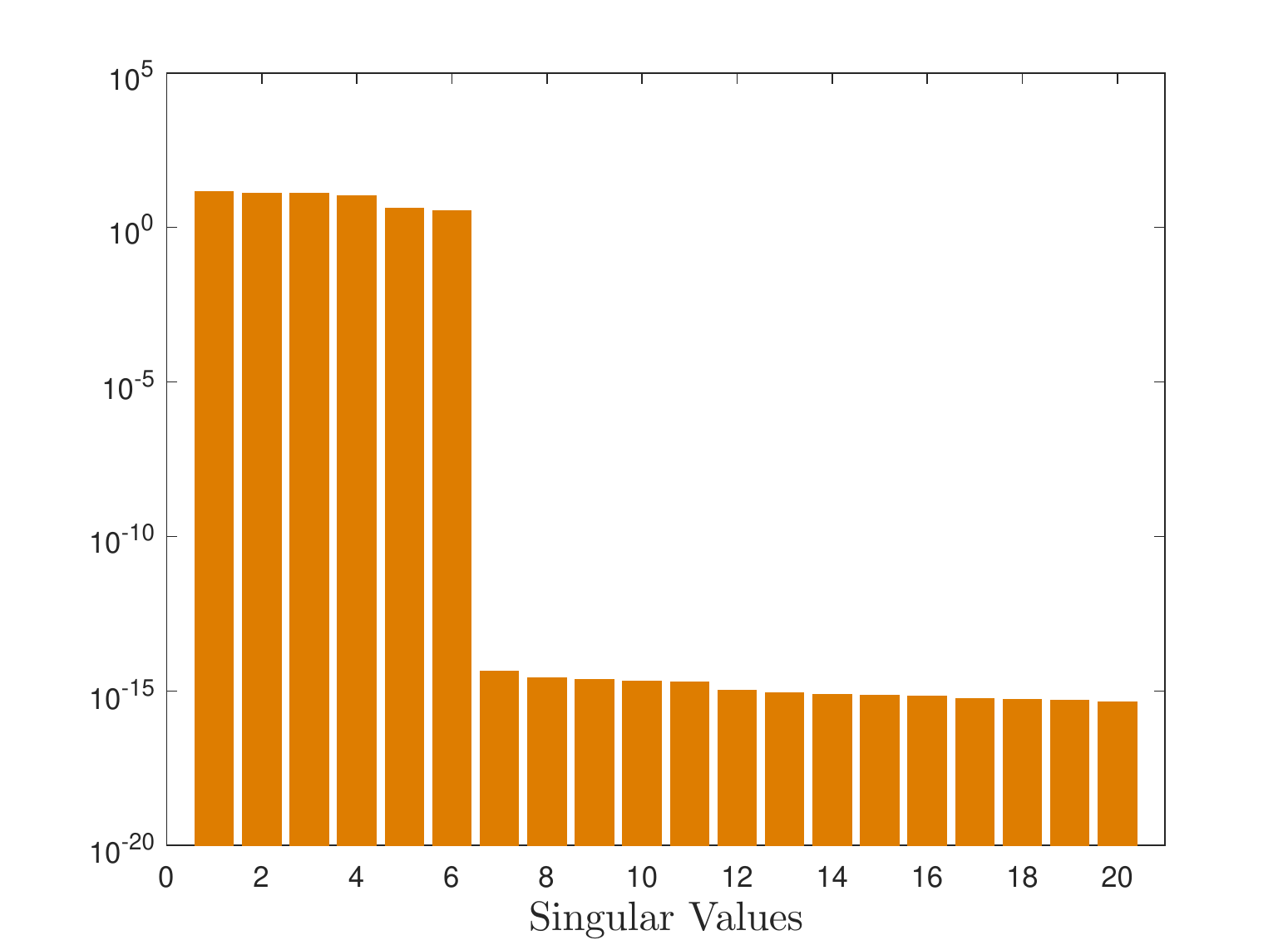}}
	\caption{Singular values of $\bm{H}_{rs}(0)$.}
	\label{singu2}
\end{figure}

Moreover, with the sampling time $\Delta t=0.1\mu s$ and the final time $T=12\mu s$, the measurement and the real results of the output $\langle\sigma_x^1\rangle$ are plotted as the black-dot and yellow-dot lines, respectively, in Fig. \ref{output}, where the polluted measurement result has a discrepancy of its real value. However, the measurement result $\{\tilde{y}(k)\}$ can still be used to construct the Hankel matrix $\bm{H}_{rs}(0)_{rL\times s}$ where we let $r=20, L=1, s=100$. Consequently, we can obtain the corresponding singular value decomposition, where we plot the singular values of $\bm{H}_{rs}(0)$ in logarithmic scale in Fig. \ref{singu2}. We can easily find a huge gap between the dominant singular values and the other quite small ones. Therefore, we can determine the dimension of the augmented system (\ref{system}) according to the number of the dominant singular values and thus the order of the colored noise realization can be determined. Hence, we have the dimension of the augmented system $\check{n}=6$ and the order of the noise realization $n=2$.


\begin{figure}[h]
	\centerline{\includegraphics[width=3in]{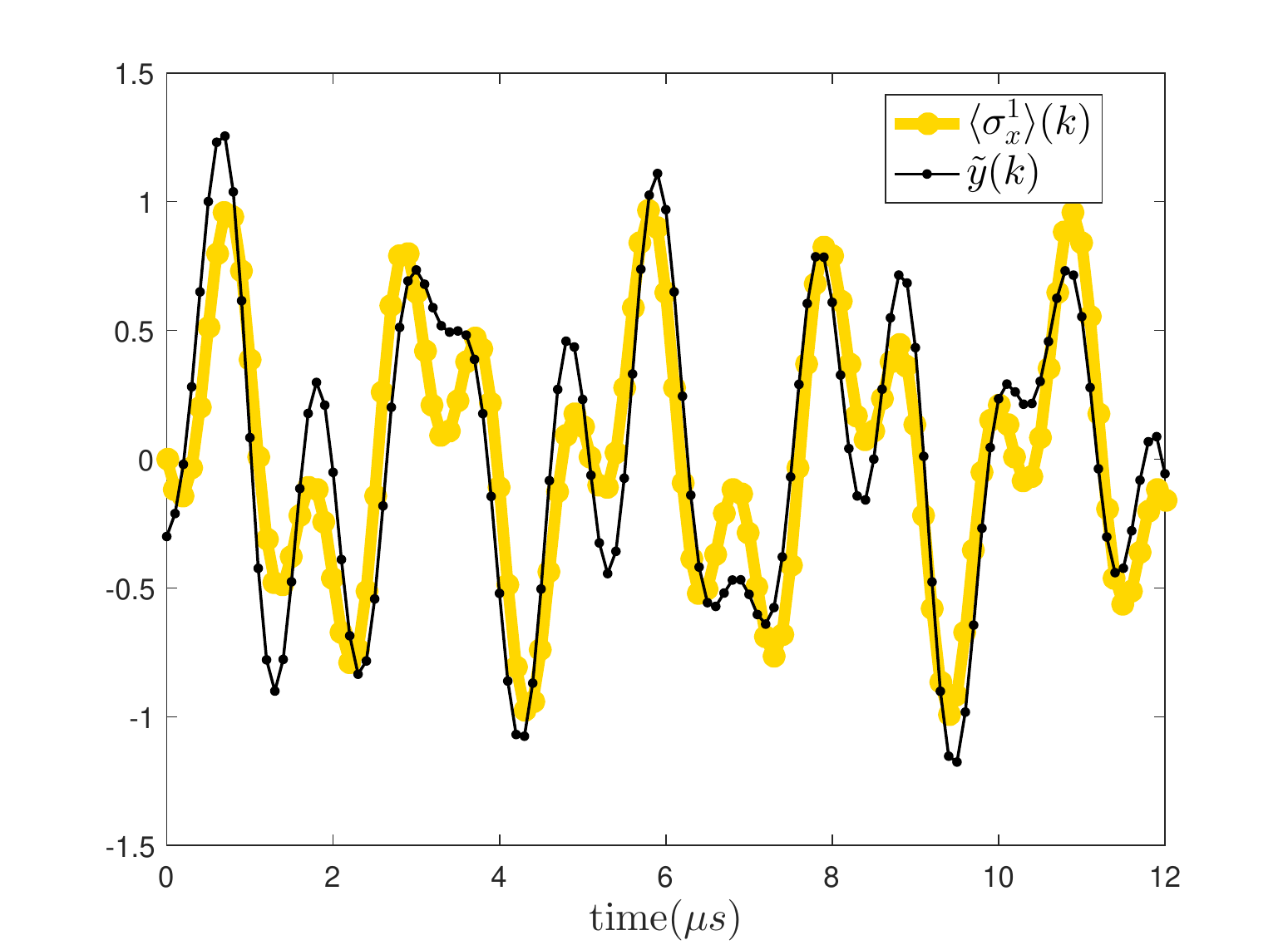}}
	\caption{Real evolution of the expectation value of $\langle\bm{\sigma}_x\rangle$ and the measured value disturbed by the classical colored noise.}
	\label{output}
\end{figure}

Following the procedure in Section~\ref{identification method} of the ERA, a realization $(\bm{\hat{A}},\bm{\hat{C}},\bm{\hat{x}}(0))$ with the dimension $\check{n}$  can be identical to that of $(\bm{\check{A}},\bm{\check{C}},\bm{\check{x}}_0)$ (\ref{system}); i.e.,
\begin{equation}\label{transformEqual}
  \bm{\check{C}}(s\bm{I}-\bm{\check{A}})^{-1}\bm{\check{x}}_0=\bm{\hat{C}}(s\bm{I}-\bm{\hat{A}})^{-1}\bm{\hat{x}}_0.
\end{equation}
whose both sides are polynomials in $s$. Specifically, the left-hand side of (\ref{transformEqual}) contains the parameters to be identified while the right-hand side of (\ref{transformEqual}) is numerically constructed by the  measurement data. Equalling the coefficients of $s$ in the same order on the both sides, we can collect a polynomial equation set. In our example, the nine lowest order polynomial equations containing nine unknown variables are
\begin{align*}\label{poly1}
     -0.3=& \xi_{01}+\xi_{02} \\
     -0.1=& e_{11}+e_{22} \\
     -0.48=& \omega_1 + e_{11}\xi_{02} - e_{12}\xi_{02}-e_{21}\xi_{01} +e_{22}\xi_{01} \\
     -13.459=&e_{11}\omega_1+ e_{22}\omega_1+2\delta_1^2\xi_{01}+ 2\delta_1^2\xi_{02}\\
     &+\omega_1^2\xi_{01}+\omega_1^2\xi_{02}+\omega_2^2\xi_{01}+\omega_2^2\xi_{02}\\
     -89.9734=& \omega_1\omega_2^2-\delta^2\omega_2+e_{11}\omega_1^2\xi_{02}+e_{11}\omega_2^2\xi_{02}\\
     &-e_{12}\omega_1^2\xi_{02}-e_{12}\omega_2^2\xi_{02}-e_{21}\omega_1^2\xi_{01}\\
     &+e_{22}\omega_1^2\xi_{01}+e_{22}\omega_2^2\xi_{01}+e_{11}e_{22}\omega_1\\
     &-e_{12}e_{21}\omega_1+2e_{11}\delta_1^2\xi_{02}-2e_{12}\delta_1^2\xi_{02}\\
     &-2e_{21}\delta_1^2\xi_{01}+2e_{22}\delta_1^2\xi_{01}\\
     67.182=& e_{11}\delta_1^2\omega_2-\delta_1^4\xi_{02}-\omega_1^2\omega_2^2\xi_{01}-\omega_1^2\omega_2^2\xi_{02}\\
     &-\delta_1^4\xi_{01}+e_{22}\delta_1\omega_2-e_{11}\omega_1\omega_2^2\\
     &-e_{22}\omega_1\omega_2^2+2\delta_1^2\omega_1\omega_2\xi_{01}+2\delta_1^2\omega_1\omega_2\xi_{02}\\
     64.43=&2\delta_1^2+\omega_1^2+\omega_2^2+e_{11}e_{22}-e_{12}e_{21}\\
     -4.443=&2e_{11}\delta_1^2+2e_{22}\delta_1^2-e_{11}\omega_1^2+e_{11}\omega_2^2\\
     &+e_{22}\omega_1^2+e_{22}\omega_2^2\\
     1124.837=&\delta_1^4+\omega_1^2\omega_2^2-2\delta_1^2\omega_1\omega_2+2e_{11}e_{22}\delta_1^2\\
     &-2e_{12}e_{21}\delta_1^2+e_{11}e_{22}\omega_1^2-e_{12}e_{21}\omega_1^2\\
     &+e_{11}e_{22}\omega_2^2-e_{12}e_{21}\omega_2^2
\end{align*}
where $e_{ij}$ is the element of $\bm{E}$ in the $i$-th row and $j$-th column and $\xi_{0i}$ is the $i$-th element of $\bm{\xi}_0$. A set of solutions can be obtained utilizing a PHCpack \cite{journals/toms/Verschelde99}.

\begin{table}[htbp]
\caption{Comparison of the solutions obtained by the ZS and our methods}\label{table1}
\begin{center}
\setlength{\tabcolsep}{0.5mm}{
\begin{tabular}{cccc}
\toprule
& $\bm{\omega_1(\text{GHz})}$ & $\bm{\omega_2(\text{GHz})}$ & $\bm{\delta_1(\text{GHz})}$\\
\midrule
\textbf{real values} & $1.3$ & $2.4$ & $4.3$\\
\midrule
\textbf{ZS method in~\cite{PhysRevLett.113.080401}} & $0.8746$ & $2.0601$ & $4.1211$\\
\midrule
\textbf{our method}& $1.3$ & $2.4$ & $4.3$\\
\bottomrule
\end{tabular}
}
\end{center}
\end{table}

In Table \ref{table1}, we compare our identified result with that obtained using Zhang and Sarovar's method (ZS method)~\cite{PhysRevLett.113.080401} which is not specially designed for the colored noise case. Our method can precisely identify the real values of the parameters in the Hamiltonian if the number of copies is not limited. While the results obtained by the method in~\cite{PhysRevLett.113.080401} are different from the real values.

\begin{figure}[htbp]
\centerline{\includegraphics[width=3in]{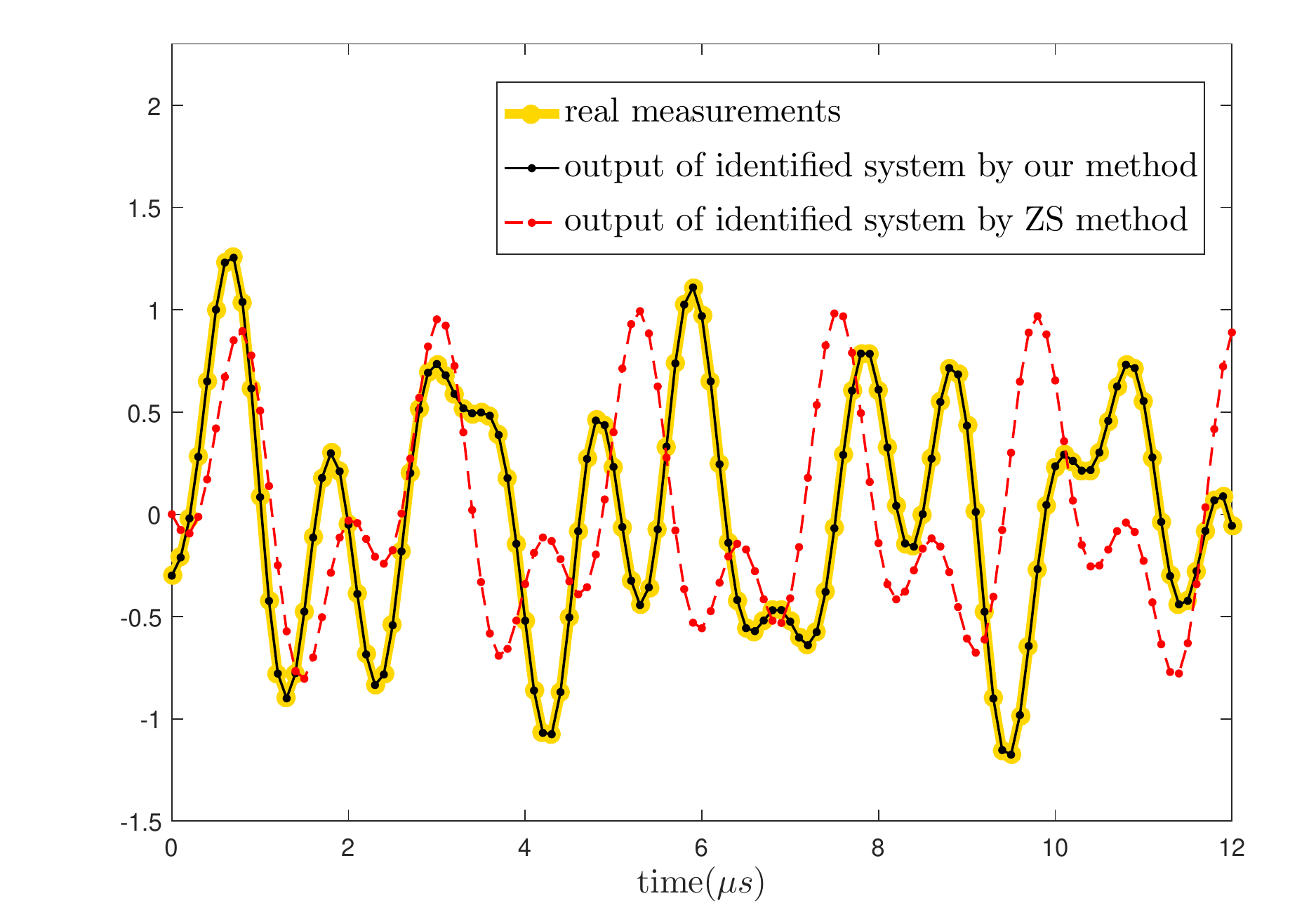}}
\caption{Comparison between real measurements and outputs of identified systems obtained from our method and ZS method.}
\label{identified_output}
\end{figure}

Moreover, in our method the estimates for the noise realization are $e_{11}=10^6(-50.025+410.92i), e_{12}=10^6(-49.97+415.41i), e_{21}=10^6(49.94-406.43i), e_{22}=10^6(49.925-410.92i), \xi_{01}=10^6(-4053.8+3692.7i), \xi_{02}=10^6(4053.5-410.92i)$.  After substituting the solution of our method back into the augmented system model (\ref{system}) and the solution of ZS method back into the quantum system model (\ref{realization2}), the outputs produced by the two identified systems are compared with the real measurements in Fig. \ref{identified_output}. It can be seen that our identification results coincide with the real system which shows that our method can improve the identification accuracy with classical colored measurement noise.

\section{Conclusion}\label{conclu}
In this paper, we designed a  procedure to identify the Hamiltonian of closed quantum system under classical colored measurement noise and showed its performance on a two-qubit example. In our method, an augmented system model was constructed and ERA was used to eliminate the impact of classical colored measurement noise on the precision of Hamiltonian identification. In principle, our identification procedure is applicable to various colored noise in the measurement process and does not require any prior information (e.g., PSD) about noise. The future research will focus on extending our method to open quantum systems for Hamiltonian identification when there exists classical colored measurement noise.

\bibliographystyle{IEEEtran}
\bibliography{Hamiltonianidentificationcolorednoise}

\end{document}